\documentstyle[12pt]{article}

\newcommand{\LL}{{I\!\! L}}
\newcommand{\half}{\frac{1}{2}}
\newcommand{\vx}{{\bf x}}
\newcommand{\vy}{{\bf y}}
\newcommand{\eq}[1]{(\ref{#1})}
\newcommand{\diff}{\partial}
\newcommand{\beq}{\begin{equation}}
\newcommand{\eeq}{\end{equation}}
\newcommand{\beqn}{\begin{eqnarray}}
\newcommand{\eeqn}{\end{eqnarray}}
\newcommand{\ddd}{{\rm d}}
\newcommand{\cR}{{\cal R}}
\newcommand{\cD}{{\cal D}}
\newcommand{\dddd}{\cD}
\newcommand\Appendix[1]{\par
 \setcounter{section}{0}
 \setcounter{equation}{0}
 \renewcommand{\thesection}{Appendix \Alph{section}}
 \section{#1}
 \def\theequation{\Alph{section}.\arabic{equation}}}
\def\cF{{\cal F}}
\def\cC{{\cal C}}
\def\cW{{\cal W}}
\newcommand{\cZ}{{\cal Z}}
\newcommand{\cS}{{\cal S}}
\def\cL{{I\!\! L}}
\def\dd{{\rm d}}
\def\NP{ Nucl.~Phys.}
\def\PR{ Phys.~Rev.}
\def\PL{ Phys.~Lett.}
\def\PRL{ Phys.~Rev.~Lett.}

\title{
Quantum Theory of Strings in Abelian Higgs Model}

\author{ E.T.~Akhmedov$^{1,2}$,
M.N.~Chernodub$^{1,2}$,
M.I.~Polikarpov$^{1}$\\ and M.A.~Zubkov$^{1}$\\
{\sl $^1$Institute of Theoretical and Experimental
Physics,}\\ {\sl B.Cheremushkinskaya 25, Moscow, 117259 Russia.}\\
{\sl $^2$Moscow Institute of Physics and Technology,}\\ {\sl Dolgoprudny,
Moscow region, Russia.}
}

\date{}
\begin{document}
\bibliographystyle{bibstand}
\maketitle
\begin{abstract}
Starting from the Abelian Higgs field theory, we construct the theory of
quantum Abrikosov--Nielsen--Olesen strings. It is shown that in four space
-- time dimensions in the limit of infinitely thin strings, the conformal
anomaly is absent, and the quantum theory exists. We also study an analogue
of the Aharonov--Bohm effect: the corresponding topological interaction is
proportional to the linking number of the string world sheet and the
particle world trajectory. The creation operators of the strings are
explicitly constructed in the path integral and in the Hamiltonian
formulation of the theory. We show that the Aharonov--Bohm effect gives rise
to several nontrivial commutation relations.
\end{abstract}

\newpage

\section{Introduction}
One of the principal problems of the quantum field theory is the search of
the vacuum $\Psi$ function.  This problem is especially important for the
nonperturbative description of the gluodynamics and chromodynamics.  The
standard way to obtain the nonperturbative effects is to use some vacuum
consisting of instanton like classical solutions. In the present publication
we consider the vacuum consisting of Abrikosov--Nielsen--Olesen (ANO)
strings \cite{AbNiOl}. We start from the quantum Abelian Higgs theory, in
which ANO strings are classical solutions.  This theory can be considered as
a relativistic generalization of the effective theory of the superconductor
near the critical point (Ginsburg-Landau theory), and we do not pay
attention to the zero-charge problem. We work in the Euclidian space and,
taking into account the measure, extract from the functional integral the
part corresponding to the topological defects which are ANO strings.  We can
perform all calculations for the case when the world sheets have the
topology of the sphere.  Actually, we perform in the continuum limit the
same transformations that have been used in the lattice compact QED
\cite{BaMyKo77} and in the lattice Abelian Higgs model \cite{PoWiZu93}.  It
was shown that the partition function for the compact fields on the lattice
can be factorized:  $\cZ_{com} = \cZ_{ncom} \cdot \cZ_{top}$, where
$\cZ_{ncom}$ is the partition function for the noncompact fields, and
$\cZ_{top}$ is the partition function for the topological
defects\footnote{The transformation of this type have been considered for
the first time for the two-dimensional lattice $XY$ model. It was shown
\cite{Ber70,KoTh73} that the partition function of the $XY$-model is
equivalent to the partition function of the Coulomb gas. For the three- and
the four-dimensional $XY$ model it is also possible \cite{BuPoPoWi93} to get
the partition function for the topological defects, which are vortex lines
and ``global strings''.} (monopoles in compact QED and strings in the
Abelian Higgs model.).

In first papers on the quantum ANO strings \cite{For74}, \cite{GeSa75},
where the London limit (infinitely massive Higgs boson) was considered, it
was shown that in the strong coupling limit (thin and long strings) the
strings can be described by the Nambu-Goto action. The exact action for the
ANO strings in the London limit is obtained in \cite{Orl94}. It was
shown that in the string action there are terms depending on the
powers of tensor of extrinsic curvature with exponents $>2$. These
terms ensure the stability of the classical string. The tree level
corrections to the ANO string action were studied in \cite{SaYa94}. The
duality transformation for the Abelian Higgs model was discussed in
\cite{Lee93}.

It is impossible to get the quantum theory of the ANO strings from the
actions discussed in \cite{For74}--\cite{Lee93}. If we consider the limit of
infinitely thin strings, the theory becomes conformal, and it is well
known that there are difficulties with the quantization of this theory in
4D: there either exists the conformal anomaly \cite{Pol81} (in the case of
Hamiltonian or path integral quantization) or Lorentz invariance is absent
(in the case of the light cone quantization). So we start from the quantum
Abelian Higgs field theory, and it seems that we get the string theory which
cannot be quantized. As shown below, the solution of this paradox lies in
the accurate change of the field variables to the string variables.

An example of an effective theory of infinitely thin quantum 4D ANO
strings was suggested by Polchinski and Strominger \cite{PoSt91}. It
occurs that if one adds to the Nambu-Goto action an additional term, then
this theory can be quantized in the Hamiltonian formalism. For the
Nambu-Goto action we have, on the classical level, the Virassoro algebra
(algebra of generators of conformal transformations):

\beq
[L^{NG}_{n}, L^{NG}_{m}] = (n - m) L^{NG}_{n + m}.
\eeq
On the quantum level, taking into account the raparametrization ghosts, we
have for the pure Nambu-Goto action:

\beqn
[\cL^{NG}_{n}, \cL^{NG}_{m}] & = & (n - m) \cL^{NG}_{n + m} + \frac{D -
26}{12} (m^3 - m) \delta_{n + m, 0}, \\ \qquad
\cL^{NG}_{n} & = & L^{NG}_{n} + L^{gh}_{n},  \nonumber
\eeqn
where $D$ is the dimension of the space-time and $L^{gh}_{n}$ are Virassoro
generators which arise due to the ghost fields. If
we add the term suggested in \cite{PoSt91} with an arbitrary
coefficient $\gamma$ to the string action, Virassoro algebra for the full
generators $\cL^{tot}_{n} = \cL^{NG}_{n} + L^{\gamma}_{n}$ takes the
form\footnote{The algebra \eq{VAfinal} was obtained in \cite{PoSt91} in
the leading order of the expansion in ${\bar R}^{-1}$, where ${\bar R}$ is
the mean curvature of the strings.}:

\beqn
[\cL^{tot}_{n}, \cL^{tot}_{m}] = (n - m) \cL^{tot}_{n + m} + \frac{D -
26 + \gamma}{12} (m^3 - m) \delta_{n + m, 0}. \label{VAfinal}
\eeqn
Therefore, by adjusting $\gamma$, we can cancel the conformal anomaly for
$D=4$ \cite{PoSt91}.

In Section 2 we show that such additional term in the action naturally
arises for the ANO strings, if we take into account the Jacobian of the
transformation from the field variables to the string variables.
A preliminary, not although completely correct, calculation of the
this Jacobian is published in \cite{AkhZu95}. The usual terms are also
present. The first two terms in the expansion of the action of the ANO
strings \cite{Orl94,SaYa94}, in powers of the average inverse string
curvature, are the standard term proportional to the area of the string
world sheet, and the rigidity term \cite{Pol86,Kle86} with negative
sign.

In the refs. \cite{AlWi89,AlMaWi90,PrKr90,KrWi89,BaMoPr93} the topological
long--range interaction of the strings and charged particles was discussed.
This interaction was discussed in \cite{PoWiZu93} for the string
representation of the $4D$ {\sl lattice} Abelian Higgs model. In Section 3
we repeat the calculations of \cite{PoWiZu93} in the continuum limit and
show explicitly the existence of the Aharonov--Bohm effect in the field
theory. The reason for this long--range interaction is that the charges
$M = e,\ 2e, \ \ldots (N-1)e$ cannot be completely screened by the
condensate of the field of charge $Ne$; if $M/N$ is integer, then the
screening is complete and there are no long--range forces.

In Section 4 we construct the operator which creates the
string in a given time slice on the contour $\cC$. This operator is the
continuum analogue of the lattice operator considered in
\cite{BuPoPoWi93,PoWiZu93}.

In Section 5 we consider the theory in the Hamiltonian formalism and
show that the string creation operator has nontrivial commutation
relations\footnote{By a nontrivial commutation relation we mean a
relation of the type: $A B - e^{i\xi} B A=0$.} with the Wilson loop
operator; this is a direct consequence of the Aharonov--Bohm effect. We
give several other examples of nontrivial commutation relations.

\section{From the Abelian Higgs model to the Quantum Strings}

The partition function for the four-dimensional Abelian Higgs Model is given
by the formula:

\beqn
\cZ = \int \cD A_{\mu} \cD \Phi \exp \left\{ - \int \ddd^4 x \left[
\frac{1}{4} F^2_{\mu \nu} + \frac{1}{2} |D_{\mu} \Phi|^2 + \lambda
(|\Phi|^2 - \eta^2)^2 \right] \right\}, \\
\qquad D_{\mu} = \diff_{\mu} - i e A_{\mu}\,, \label{Initial}
\eeqn
throughout the paper we assume the Pauli--Villars regularization, we discuss
some details of the regularization in the Appendix.

In the equation \eq{Initial} the integration
over the complex scalar field $\Phi = |\Phi| \exp{(i \theta)}$ can be
rewritten as:

\beq
\int \cD \Phi... = \int \cD Re \Phi \cdot \cD Im \Phi... = \int
[|\Phi| \cD |\Phi|] \cD \theta... \, .  \label{PP}
\eeq
The functional integral over $\theta$ should be carefully defined, since
$\theta$ is not defined on the manifolds where

\beq
 Im \Phi = Re \Phi = 0\,. \label{ImRe}
\eeq
These two equations define the two--dimensional manifolds in the four
dimensional space--time and we should integrate over all functions that
are regular everywhere except for these two-dimensional manifolds.  These
two-dimensional singularities are nothing but the ANO string world sheets,
since the Higgs field is zero at the center of the ANO string.

In eq. \eq{PP} we integrate over the regular functions $Re \Phi(x)$ and $Im
\Phi(x)$, and it can be shown \cite{Orl94,SaYa94,Lee93,BaSa78} that the
singularities in the function $\theta(x)$ defined by eqs.\eq{ImRe} should
have the form:

\beqn
\diff_{[\mu,} \diff_{\nu]} \theta^s (x, \tilde x) & = &
2 \pi \epsilon_{\mu\nu \alpha \beta}
\Sigma_{\alpha \beta}(x, \tilde x), \label{theta}\\
\Sigma_{\alpha \beta}(x, \tilde x) & = &
\int_{\Sigma} \ddd \sigma_{\alpha \beta}(\tilde x) \delta^{(4)}
[x -\tilde x(\sigma)], \nonumber\\
\ddd \sigma_{\alpha \beta}(\tilde x) & = & \epsilon^{ab}
\diff_a \tilde x_{\alpha} \diff_b \tilde x_{\beta} \ddd^2 \sigma = \sqrt{g}
t_{\alpha \beta} \ddd^2 \sigma, \nonumber
\eeqn
where $\tilde x = \tilde x(\sigma)$ are the coordinates of the
two-dimensional singularities parametrized by $\sigma_a, \ a = 1, 2$;
$\diff_a = \frac{\diff}{\diff \sigma_a}$; $\theta^s (x, \tilde x)$ is the
function of $x$ and the functional of $\tilde x$; $\Sigma$ defines the
position of the singularities; $g = det||g_{ab}||$; $g_{ab} = \diff_a \tilde
x_{\mu} \diff_b \tilde x_{\mu}$ and $t_{\mu \nu} =
\frac{\epsilon_{ab}}{\sqrt{g}} \diff_a \tilde x_{\mu} \diff_b \tilde
x_{\nu}$ are the tensors of the induced metric and the extrinsic curvature
on $\Sigma$ (we have no intrinsic metric in the theory), $t^2_{\mu \nu} =
2$. Note that $\diff_{[\mu} \diff_{\nu]} \theta^s \neq 0$ since
$\theta^s$ is a singular function.

For simplicity, we consider the London limit ($\lambda \rightarrow \infty$
and the radial part of the field $\Phi$ is fixed)\footnote{For an
arbitrary $\lambda$ all the transformations remain the same, but in the
final expression for the partition function \eq{ii} we have an additional
functional integral over the radial part of the field $\Phi$.}:

\beqn
\cZ = const \cdot \int \cD A_{\mu} \cD \theta \exp \left\{ - \int
\ddd^4 x \left[ \frac{1}{4} F^2_{\mu \nu} + \frac{\eta^2}{2}
|\diff_{\mu} \theta + e A_{\mu}|^2 \right] \right\}, \label{zzz}
\eeqn
where $\eta^2 = <|\Phi|^2>$. Now we discuss the measure of the integration
over $\theta$. From \eq{PP} it follows that the norm for the field $\theta$
is: $||\delta\theta||^2 = \int\ddd^4 x |\Phi|^2 {(\delta\theta)}^2$. In the
London limit  there are two independent variables: the regular and the
singular part of $\theta$, $\theta = \theta^{r} + \theta^s$, and
$||\delta\theta||^2 = const \int\ddd^4 x {\left( \delta \theta^r + \delta
\theta^s \right)}^2 = ||\delta\theta^r||^2 + ||\delta\theta^s||^2$. From
eqs.\eq{theta} it can be easily seen that the interference term $\int\ddd^4
x \, \delta \theta^r \delta \theta^s$ vanishes:

\beqn
 \int\ddd^4 x \delta \theta^r \delta \theta^s = const \int\ddd^4 x
 \int\ddd^4 y \delta \theta^r \diff_{[\mu} \diff_{\nu]}
 \Delta^{-1}(x-y) \delta \Sigma_{\mu\nu} = \nonumber\\
 const \int\ddd^4 x
 \int\ddd^4 y \left( \diff_{[\mu} \diff_{\nu]} \delta \theta^r \right)
 \Delta^{-1}(x-y) \delta \Sigma_{\mu\nu} = 0\,,
\eeqn
we use the fact that $\diff_{[\mu} \diff_{\nu]} \delta\theta^r = 0$, since
$\delta\theta^r$ is a regular function. Therefore, $\int \cD \theta...  =
\int \cD \theta^{r} \cD \theta^s...$, and now we can show that the integral
over the singular part $\theta^s$ can be reduced to the integral over the
string world sheets. We have no monopoles in the theory; therefore, due to
the conservation of the magnetic flux, the ANO strings
are closed, and the singularities, defined by $\theta^s$ ($\Sigma$ in
eq.\eq{theta}), form closed two-dimensional surfaces.  In the
infinite space--time $\cR^4$ the strings which are closed through the
boundary conditions have the infinite action, therefore we do not take them
into account.

Now, let us transform the partition function of the field theory \eq{zzz} to
the partition function of the string theory. In order to change the
integration variables, we substitute the unity into the functional integral
\eq{zzz} (see eq. \eq{theta}):

\beqn
1 = \tilde J [\Sigma_{\mu \nu}] \cdot \int \cD \tilde x
\cdot \delta \left\{ \Sigma_{\mu \nu} -
\int_{\Sigma} \ddd^2 \sigma \sqrt{g} t_{\mu \nu}
\delta^{(4)} [x - \tilde x(\sigma)] \right\}.  \label{def}
\eeqn
Here $\tilde J [\Sigma_{\mu \nu}]$ is the Jacobian which corresponds to the
change of the field variables to the string variables, and in $\int \cD
\tilde x$ we assume summation over the topologies of the string world
sheets. Using the $\delta$--function in \eq{def} and the definition of
$\theta^s$ \eq{theta}, we integrate over $\theta^s$ in the partition
function:

\beqn
\int \cD \theta e^{-S[\theta,...]} =
\int \cD \theta^{r} \cD \theta^s e^{-S[\theta^r+\theta^s, ...]} =
\label{kk}\\ const \cdot \int \cD \theta^{r} \cD \Sigma_{\mu\nu}
e^{-S[\theta^r +\theta^s(\Sigma_{\mu\nu}), ...]} =
const \cdot \int \cD \theta^{r} \cD \tilde x
J(\tilde x) e^{-S[\theta^r + \theta^s(x,\tilde x), ...]} \,, \nonumber
\eeqn
where $J(\tilde x) = \tilde J [\Sigma_{\mu \nu}]$.

Fixing the gauge $\diff_{\mu} \theta^{r} = 0$, it is easy to perform
integration over $A_{\mu}$; the result is:

\beqn
\cZ = \int \cD \tilde x J(\tilde x)
\cdot \exp \left\{ - \eta^2 \pi^2 \int_{\Sigma}
\int_{\Sigma} \ddd \sigma_{\mu \nu}(\tilde x)
\cD_m^{(4)}(\tilde x - \tilde x') \ddd \sigma_{\mu \nu}(\tilde x')
\right\}, \label{ii}
\eeqn
where $(\Delta + m^2) \cD_m^{(4)}(x) = \delta^{(4)}(x)$, and $m^2 = e^2
\eta^2$ is the mass of the gauge boson. The action which enters the
partition function \eq{ii} was already discussed in
\cite{Orl94,SaYa94}, the new object in \eq{ii} is the Jacobian
$J(\tilde x)$.
It easy to see that $J(\tilde x)$  defined by \eq{def} and the resulting
partition function \eq{ii} are invariant under the raparametrization of the
coordinate $\sigma$ on the world sheet. As shown in the Appendix,
$J(\tilde x)$ can be evaluated if the string world sheet $\Sigma$ has the
spherical topology. The calculations are performed in the conformal gauge,

\beq
g_{12}=g_{21}=0\,;\quad g_{11}=g_{22}=\sqrt{g}\, ,
\label{vv}
\eeq
and the result is:

\beqn
J(\tilde x) = const \cdot \exp \left\{ \int_{\Sigma} \ddd^2 \sigma
\left[\frac{11}{48 \pi} (\diff_a \ln\sqrt{g})^2  + \mu_1
\sqrt{g} + \frac{\mu^2_3\ln{\Lambda_1 \bar R}}{\mu^2_2\pi}
\sqrt{g} (\diff_a t_{\mu \nu})^2 \right]\right\}\, ,  \label{YY}
\eeqn
the parameters $\mu_i$ are defined in the
Appendix, $\Lambda_1$ is the regularization parameter, and $\bar R$ is the
average curvature of the strings.

Now we study the expansion of the action in powers of ${(m \bar R)}^{-1}$.
A similar expansion was studied in refs. \cite{Orl94,SaYa94}, but we
include in the expansion the terms which come from $\ln J(\tilde x)$. In
the leading order, the action is local and, if the surface $\Sigma$ has the
spherical topology, we use the expression (\ref{YY}) for $J$; the result
is:

\beqn
S = \mu' \int_{\Sigma} \ddd^2 \sigma \sqrt{g} - \frac{11}{48 \pi}
\int_{\Sigma} \ddd^2 \sigma (\diff_a \ln{\sqrt{g}})^2 - \beta \int_{\Sigma}
\ddd^2 \sigma \sqrt{g} (\diff_a t_{\mu \nu})^2\,. \label{ee}
\eeqn
Here $ \mu' = \mu_0 - \mu_1$, the string tension
$\mu_0$ comes from the expansion of the action \eq{ii}
\cite{Orl94,SaYa94}, in the regularization scheme accepted in \cite{Orl94}
$\mu_0 = 4 \pi \eta^2 \ln{(\frac{\Lambda^2_2}{m^2})}$; $\mu_0$ is
renormalized by $\mu_1$ which enters the Jacobian \eq{YY}.  $\beta=
\frac{\pi}{4 e^2} + \frac{\mu^2_3\ln{\Lambda_1 \bar R}}{\mu^2_2\pi}$ where
the first term in the r.h.s comes from the expansion of the action, the
second one is due to the Jacobian.

The first term in \eq{ee} is the usual Nambu--Goto action;
the second term, as we said, is important for the quantization, and the
third one is the rigidity term (see \cite{Pol86,Kle86}).

If we consider the strings without rigidity, $\beta = 0$,
we get the theory studied in \cite{PoSt91}. It occurs that the
coefficient of the second term in \eq{ee} corresponds to $\gamma = 22$ in
the Virassoro algebra \eq{VAfinal}; therefore, the conformal anomaly is
absent and the theory can be quantized in $D=4$. It should be emphasized
that this term appears from the Jacobian $J(\tilde x)$.

It is obvious from the derivation of $J(\tilde x)$ that it has universal
nature, i.e. it is independent on the model under consideration. The
Jacobian $J(\tilde x)$ arises when we pass from the integration over the
field variables to the integration over the string variables. Therefore, we
expect that any field theory which has the string-like solutions is
equivalent to the string theory which can be quantized in $D = 4$. As
mentioned in ref.~\cite{Orl94}, the action which enters the partition
function \eq{ii} leads to the stable ANO strings, but the
dominant vacuum configuration is branched polymers formed by
the string world sheets. It would be interesting to study the dominant
vacuum configuration of the strings, taking into account the Jacobian
$J(\tilde x)$.

\section{The Aharonov--Bohm Effect in the Abelian Higgs Model.}

Now we consider the Abelian Higgs model with the Higgs bosons
carrying the charge $N e$, the partition function now is:

\beq
 \cZ = \int \cD A_{\mu} \cD \theta
 \exp \left\{- \int \ddd^4 x \left [ \frac{1}{4} F^2_{\mu \nu} +
 \frac{\eta^2}{2} {\left( \diff_{\mu} \theta + N e A_{\mu} \right)}^2
 \right] \right \}\,. \label{AHMN}
\eeq
There exists a nontrivial long--range topological interaction of
Nielsen--Olesen strings with particles of charge $Me$, if $\frac MN$ is
noninteger. This is the four--dimensional analogue
\cite{AlWi89,AlMaWi90,PrKr90} of the Aharonov--Bohm effect studied for the
lattice Abelian Higgs model in \cite{PoWiZu93}.  Now we derive the
long range interaction, using the string representation of the theory.
Consider the Wilson loop for the particle of the charge $M e$:

\beq
   W_M(\cC) = \exp\left\{ i M e \int \ddd^4 x j^\cC_\mu (x) A_{\mu}(x)
   \right\} = \exp\left\{ i M e \int_{\cC} \ddd x^{\mu} A_{\mu}(x)
   \right\}\,, \label{WM}
\eeq
where the current is the $\delta$--function on a contour $\cC$:

\beqn
j^{\cC}_\mu(x) = \int_{\cC} \ddd t \dot{\tilde z}^{\mu}(t) \delta^{(4)}(x -
\tilde z(t)) \label{jC}
\eeqn
and the function $\tilde z_{\mu}(t)$ parametrizes the contour.

Substituting \eq{WM} into the path integral \eq{AHMN} and changing the
field variables to the string variables, as described in the previous
section, we obtain:

\beqn
<W_M(\cC)> =  \nonumber \\
  \frac{1}{\cZ} \int \cD \tilde x J(\tilde x)\exp \Biggl\{
 -  \int \ddd^4 x \int
 \ddd^4 y \biggl [
 \pi^2 \eta^2 \Sigma_{\mu\nu}(x)
 \cD_m^{(4)}(x - y) \Sigma_{\mu \nu}(y)
 \nonumber \\
 + \frac{M^2 e^2}{2} j^{\cC}_\mu(x) \cD^{(4)}_m(x - y)
 j^\cC_\mu(y) + \pi i \frac{M}{N} j^\cC_\mu(x)
 \cD^{(4)}_m(x - y) \diff_{\nu} \epsilon_{\mu \nu \alpha \beta}
 \Sigma_{\alpha \beta}(y)\biggr ] \nonumber \\
 + 2 \pi i \frac{M}{N}  \LL(\Sigma,\cC)
 \Biggr\}\,, \label{bb}
\eeqn
where  $m = Ne\eta$ is the boson mass, and

\beqn
 \LL(\Sigma,\cC) = \half \int \ddd^4 x \int
 \ddd^4 y \epsilon_{\mu \nu \alpha \beta} \Sigma_{\mu\nu}(x)
 j_{\alpha}^{\cC}(y)  \diff_{\beta} \cD^{(4)}_0(x - y)  = \nonumber\\
 = \frac{1}{4 \pi^2}
 \int \dd^4 x \int \dd^4 y\, \epsilon_{\mu\nu\alpha\beta}\,
 \Sigma_{\mu\nu}(x)\, j^{\cC}_{\alpha}(y)\,\frac{{(x-y)}_{\beta}
 }{{|x-y|}^4} \label{Link4D}
\eeqn
is the linking number of the string world sheet $\Sigma$ and the trajectory
of the charged particle $\cC$, this formula represents a four--dimensional
analogue of the Gauss linking number for loops in three dimensions.
The first three terms in the exponent in \eq{bb} are short range
interactions and self--interactions of strings and the tested particle. The
forth term is the long--range interaction which describes the
four--dimensional analogue \cite{AlWi89,AlMaWi90,PrKr90,KrWi89} of the
Aharonov--Bohm effect:  strings correspond to solenoids which scatter
charged particles. $\LL$ is an integer, and if $M/N$ is an integer
too, then there is no long--range interaction; this situation corresponds to
such a relation between the magnetic flux in the solenoid and the charge of
the particle when the scattering of the charged particle is absent.

Another consequence of the Aharonov--Bohm effect can be obtained, if we
consider the operator $\cF_{N}(\cS)$ \cite{KrWi89} which creates the
string with the magnetic flux $\frac{2 \pi}{N e}$ moving along a fixed
closed surface $\cS$. $\cF_{N}(\cS)$ is the analogue of the Wilson loop
which creates the particle moving along the closed loop $\cC$. An
explicit form of $\cF_{N}(\cS)$ is \cite{KrWi89}:

\beq
 \cF_{N}\left(\cS\right)  =
 \exp\left\{ - \frac{\pi }{ N e}
 \int_{\cS} \dd \sigma_{\mu\nu}
  \epsilon_{\mu\nu\alpha\beta}\, F_{\alpha\beta} (x)\right\}\,.  \label{FE}
\eeq
There exists an operator which can be calculated {\it exactly},
\cite{PrKr90}; this operator is the normalized product of the Wilson
loop $W_M(\cC)$ and $\cF_{N}(\cS)$:

\beq
 A_{NM}(\cS,\cC) = \frac{ \cF_N(\cS) W_M(\cC)}{<\cF_N(\cS)> <W_M(\cC)>}.
\eeq
Here $<\cF_{N}(\cS)>$ is a constant which depends on the regularization
scheme. Substituting this operator into the functional integral \eq{AHMN}
and integrating over the fields $A$ and $\theta$, we obtain the following
result:

\beq
  <A_{NM}(\cS,\cC)> = e^{2 \pi i \frac{M}{N} \LL(\cS,\cC)}. \label{AMN}
\eeq
The meaning of this result is very simple. If the surface $\cS$ lies in a
given time slice, then $<A_{NM}(\cS,\cC)> = \exp\left\{\frac{2\pi
i}{Ne}Q_{\cS}\right\}$ (see \cite{KrWi89,PrKr90}), where $Q_\cS$ is
the total charge inside the volume bounded by the surface $\cS$; if
$\LL(\cS,\cC) = n$, then there is the charge $Mne$ in the volume bounded by
$\cS$.

\section{The String Creation Operator.}
\label{strcreopsect}

In Section 2 we have derived the partition function of the Abelian Higgs
model as a sum over the closed world sheets of the ANO strings. Now we
construct the operator which creates the string on a closed loop at a
given time; after a while the string shrinks. The vacuum expectation
value of this operator is the sum over all surfaces spanned on a given loop.
A similar operator for the lattice theory was suggested in
\cite{BuPoPoWi93,PoWiZu93}. The construction is quite the same as that
of the soliton creation operator suggested by Fr\"{o}hlich and
Marchetti \cite{FrMa87}. First we consider the model \cite{Lee93} which is
dual to the original Abelian Higgs model. It contains the gauge field
$B_\mu$ dual to the gauge field $A_\mu$, and also the hypergauge field
$h_{\mu\nu}$ dual to $\theta^r$. As in eq. \eq{kk}, we change the
integration in $\theta^s$ to the integration in $\tilde x$. The details
of the duality transformation are given in \cite{Lee93}. Taking into the
account the Jacobian, we get:

\beqn
\cZ =  \int \cD h \cD B \cD \tilde x J(\tilde x)
\exp \biggl\{ - \int \ddd^4 x \biggl[
\frac{1}{3 \eta^2} H^2_{\mu \nu \alpha} + \frac{e^2 N^2}{2} (h_{\mu
\nu} - \diff_{\mu} B_{\nu} + \diff_{\nu} B_{\mu})^2 \nonumber \samepage \\
+ 2 \pi i h_{\mu\nu} \Sigma_{\mu\nu}
(x, \tilde x) \biggr] \biggr\}\, , \label{Dual}
\eeqn
where $H_{\mu \nu \sigma} = \diff_\mu h_{\nu\sigma} + \diff_\nu h_{\sigma
\mu} + \diff_\sigma h_{\mu\nu} $ is the field strength of the
hypergauge field $h_{\mu\nu}$.
The action of the dual theory is invariant under the gauge
transformations:  $B_{\mu}(x)~\rightarrow~B_\mu(x) +
\diff_{\mu}\alpha(x)$, $h_{\mu\nu}(x)~\rightarrow~h_{\mu\nu}(x)$, and under
the hypergauge transformations: $B_{\mu}(x) \rightarrow B_{\mu}(x) -
\gamma_{\mu}(x)$, $h_{\mu\nu}(x) \rightarrow h_{\mu\nu}(x) +
\diff_{\mu}~\gamma_{\nu}(x) - \diff_{\nu}~\gamma_{\mu}(x)$.

The ANO string carries magnetic flux, and in order to construct the
creation operator, it is natural to use the dual Wilson loop: $\cW_D(\cC) =
\exp\{i \int \ddd^4 x B_{\mu}(x) j^\cC_\mu(x)\}$, where the current
$j^\cC_\mu(x)$ defines the loop $\cC$ \eq{jC}. This operator is gauge
invariant but it is not hypergauge invariant, and its vacuum
expectation value is zero. To construct the hypergauge invariant operator
\cite{BuPoPoWi93,PoWiZu93}, we follow an idea of Dirac \cite{Dir55}, who
suggested the gauge invariant creation operator of a particle with the
charge $M$:

\beq
\Phi^c_M(\vx) = \Phi_M(\vx) \exp \left \{i M e
\int \ddd^3 y G_{l}(\vx - \vy) A_{l}(\vy) \right \}\, , \label{Phic}
\eeq
here $\diff_{i} G_{i}(\vx) = \delta^{(3)}(\vx)$, and the gauge variation of
the matter field $\Phi(x) \rightarrow \Phi_M(x) \exp\{i M e \alpha(x)\}$ is
compensated by the gauge variation of cloud of photons $A_\mu$. Now we use
a similar construction, namely, we surround $\cW_D(\cC)$ by the cloud of
the Goldstone bosons:

\beq
   U(\cC) = \cW_D(\cC) \exp\left\{\frac{i}{2} \int \ddd^3 y G^{ij}_{\cC}(x -
   y) h_{i j}(y)\right\}\, . \label{Uc}
\eeq
It is easy to see that
$U(\cC)$ is hypergauge invariant if the skew--symmetric tensor
$G^{ij}_{\cC}(x)$ satisfies the equation\footnote{In this and in the next
sections, Latin indices vary from 1 to 3 and Greek ones vary from
0 to 3.} $\diff_{i} G^{ik}_{\cC}(x) = j^{\cC}_k(x)$. It is convenient to
choose $G^{ik}_{\cC}(x)$ as the surface, spanned on the loop $\cC$:
$G^{ij}_{\cC} = \int_{\cS_\cC} \ddd \sigma^{ij}(\tilde x) \delta^{(4)} [x -
\tilde x(\sigma)]$ ({\it cf.} eq.\eq{theta}). Since the string
creation operator should act at a definite
time slice, the surface defined by $G^{ij}_{\cC}(x)$ and the loop $\cC$
should belong to that time slice\footnote{The solution of the equation
$\diff_{i} G^{ik}_{\cC}(x) = j^{\cC}_k(x)$ is non--unique, moreover we
choose a two dimensional surface as the support of $G^{ik}$, the solution
which has three--dimensional support can be of the form: $G^{ik}_{\cC} =
\int \ddd^3 y \diff^{[i} j^{k]}_{\cC}(\vy) \cD^{(3)}_0(\vx - \vy) $, where
$\cD^{(3)}_0 = - \frac{1}{4 \pi |\vx-\vy|}$. It is easy to find that all
these ambiguities do not change physical results.}.

Substituting the operator \eq{Uc} into the dual partition function \eq{Dual}
and performing the inverse duality transformation,
we get the vacuum expectation value of the string creation operator in
terms of the original fields $A$ and $\theta$:

\beqn
    < U(\cC)> = \frac{1}{\cZ} \int \cD A \cD \theta
    \exp\biggl\{ - \int \ddd^4 x \biggl[ \frac{1}{4} {\left( F_{\mu\nu} +
    \frac{2 \pi}{N e} \epsilon_{\mu\nu\lambda\sigma} G^{\lambda
    \sigma}_{\cC}(x) \right)}^2
    \nonumber\\
    + \frac{\eta^2}{2} {\left( \diff_{\mu} \theta + N e A_{\mu} \right)}^2
    \biggr] \biggr\}\, , \label{UCor}
\eeqn
where the tensor $G^{\mu\nu}_{\cC}$ is equal to $G^{ij}_{\cC}$ if $\mu=i$
and $\nu=j$ are spatial indices, and $G^{0\mu}_{\cC}= G^{\mu0}_{\cC} =0$ for
any $\mu$. If we change the field variables in \eq{UCor} to the string
variables, we get a sum over closed surfaces $\Sigma$:

\beqn
<U(\cC)> = \frac{1}{\cZ} \int \cD \tilde x J(\tilde x)
\exp \left \{- \eta^2 \pi^2 \int \ddd^4 x \int \ddd^4 y \left [
\left(\Sigma^{\mu \nu}(x, \tilde x) +
\right. \right. \right. \label{<Uc>} \\
\left. \left. \left. G^{\mu \nu}_{\cC}(x) \right)
\cD^{(4)}_m(x - y) \left(\Sigma^{\mu \nu}(y, \tilde x) + G^{\mu
\nu}_{\cC}(y) \right) \right ] \right \}\,. \nonumber
\eeqn

The summation over all closed surfaces $\Sigma^{\mu \nu}$, plus the open
surface $G^{\mu \nu}$ with the boundary $\cC$, is equivalent to the
summation over all closed surfaces and over {\it all} surfaces spanned on
the loop $\cC$. Therefore, the operator $U(\cC)$ creates a string on the
loop $\cC$. Using the string creation operators, it is easy to construct the
operators which correspond to the processes of decay and scattering of the
strings.

Note that from the eq.\eq{UCor} it follows that the vacuum expectation
value $\left< U(\cC) \right>$ in the euclidean theory is positively
defined.  The fact does not mean the existence of the string condensate,
the situation is similar to the case of the Fr\"{o}hlich--Marchetti
monopole creation operator \cite{FrMa87}, see discussion in
ref.~\cite{PoPoWi91}.

If the string condensate is not zero then the infinitely large strings
contribute to the vacuum state. Formally, the string condensate exists if
in the limit $|x-y| \to \infty$:

\beq
  <U(\cC_1) \cdot U^+ (\cC_2) > \to A + B \, e^{- \mu r} + \cdots\,,
  \label{UC1UC2}
\eeq
$A \neq 0$. In eq.\eq{UC1UC2} $\cC_1$ and $\cC_2$ are finite loops at which
we create and annihilate string and $x$ ($y$) is any point on the loop
$\cC_1$ ($\cC_2$).

\section{Aharonov--Bohm Effect In The Hamiltonian Formalism.}

In this section we consider the ANO strings in the framework of the
canonical quantization. We start with the standard commutation relations:
$\left[\pi^i\left(\vx\right)\,,\, A^j\left(\vy\right)\right]= - i
\delta_{ij} \delta\left(\vx-\vy\right)$, $\pi^{i} = F^{0i}$ and
$\left[\pi_{\phi}\left(\vx\right)\,,\, \phi\left(\vy\right) \right] = - i
\delta\left(\vx - \vy\right)$, $\pi_{\phi} = {\left(D^0\phi\right)}^*$.
Using the string creation operators \eq{FE} and \eq{Uc}, we construct
several operators, which satisfy the commutator relations of the type:
$A\cdot B - B \cdot A e^{i\xi} = 0$. Similar operators are known for $3D$
Abelian models, see for example refs.\cite{D3}. The physical phenomenon
leading to the nontrivial commutation relations in the nonabelian theories
was discussed by 't~Hooft~
\cite{tHo78}.

First, let us consider the operator $U_{str}(\cC)$ which creates the
ANO string on the loop~$\cC$:

\beq
   U_{str}(\cC) = \exp\left\{ \frac{2 \pi i}{N e} \int \ddd^3 x \half
   \epsilon_{ijk} G^{ij}_{\cC}(x) \pi^{k}(x) \right\}\,, \label{Ustr}
\eeq
here $G^{ij}_{\cC}(x)$ is the same function as in
eq.\eq{Uc}. The operator \eq{Ustr} is a special case of the creation
operator:

\beq
   U[A^{cl}] = \exp\left\{ i \int \ddd^3 x
   A^{cl}_k (x) \pi^{k}(x) \right\}\,, \label{UstrA}
\eeq
where $A^{cl}(x)$ is a classical field. It is easy to see, that
$U[A^{cl}] |A(x)> = |A(x) + A^{cl}(x)>$. In \eq{Ustr} we have $A^{cl}_k (x) =
\frac{2 \pi}{N e} \epsilon_{ijk} G^{ij}_{\cC}(x)$, and the
magnetic field corresponds to the infinitely thin string on the loop $\cC$:
$B_i(x) =  \frac{2 \pi}{N e}\,  j^{\cC}_i(x)$; the current
$j_i^{\cC}$ is defined by eq. \eq{jC}.

The commutation relations for the operator \eq{Ustr} with the operators of
the electric charge $Q = \int \dd^3 x \diff_i \pi^i(\vx)$ and the magnetic
flux $\Phi_{i} = \int \dd^3 x \epsilon_{ijk} \diff^{j} A^{k}(\vx)$ also
show that $U_{str}(C)$ creates a string which carries the magnetic flux
$\frac{2 \pi}{N e}$ on the contour $\cC$:

\beqn
  \left[Q\left(x_0,\vx\right)\,,\,
  U_{str}(\cC) \right] = 0\,, \quad
  \left[\Phi^j\left(x_0,\vx\right)\,,\,
  U_{str}(\cC) \right] = \frac{2 \pi}{N e}
  j^{\cC}_i(x) U_{str}(\cC)\,, \label{commUc}
\eeqn

Note that, the string creation operator \eq{UCor} considered in the
previous section can be rewritten in the following way:

\beq
U(\cC) =
\exp\left\{ - \int \ddd^4 x \left[ \frac{1}{4} {\left( F_{\mu\nu} +
\frac{2\pi}{N e} \epsilon_{\mu\nu\lambda\sigma}
G^{\lambda \sigma}_{\cC}(x)\right)}^2 -
\frac{1}{4} F_{\mu\nu}^2\right] \right\},
\eeq
and it is clear
that, up to an inessential factor, it coincides with the definition
\eq{Ustr}.

Now we consider the commutator of the operator
$U_{str}(\cC_1)$ and the Wilson loop $W_M(\cC_2)$ \eq{WM}, the contours
$\cC_1$ and $\cC_2$ belong to the same time slice. Using the relation $e^A
e^B = e^B e^A e^{[A,B]}$, which is valid if $[A,B]$ is a $c$--number, it is
easy to get:

\beq
  U_{str}(\cC_1) W_M(\cC_2) - e^{i \xi(\cC_1,\cC_2)} W_M(\cC_2)
  U_{str}(\cC_1) = 0\,, \label{WUc}
\eeq
where
$\xi(\cC_1,\cC_2) = \frac{2 \pi M}{N} \LL(\cC_1,\cC_2)$, and
$\LL(\cC_1, \cC_2) = \frac{1}{4 \pi} \int_{\cC_1} \ddd x_i \int_{\cC_2} \ddd
y_i \epsilon^{ijk} \frac{{(x-y)}_k}{|\vx-\vy|^3}$ is the linking number of
the loops $\cC_1$ and $\cC_2$. The commutation relation \eq{WUc} is the
direct consequence of the Aharonov--Bohm effect; the wave function of the
particle of the charge $M e$ acquires the additional phase $e^{\frac{2 \pi i
M}{N}}$ if it goes around a solenoid with the magnetic flux $\frac{2
\pi}{N e}$.

The next example is the commutation relation
of the Dirac operator $\Phi^c_M(\vx)$ \eq{Phic} which creates the particle
with charge $M$ at the point $\vx$, and the operator $\cF_N(\cS)$  which
creates the string on the surface $\cS$. In Minkowsky space, the
operator $\cF_N(\cS)$ has the form (an analogue of eq.\eq{FE}):

\beq
 \cF_{N}\left(\cS\right)  =
 \exp\left\{ \frac{ i \pi }{ N e}
 \int_{\cS} \dd \sigma_{\mu\nu}
  \epsilon_{\mu\nu\alpha\beta}\, F_{\alpha\beta} (x)\right\}\,.  \label{F}
\eeq
If the surface $\cS$ belongs to the same time slice as the point $\vx$,
then:

\beq
   \cF_N(\cS) \Phi^c_M(\vx) - \Phi^c_M(\vx) \cF_N(\cS)
   e^{i\theta(\cS,\vx)} = 0\,, \label{FPh}
\eeq
where $\theta\left(\cS,\vx\right) = \frac{2 \pi M}{N}
\Theta(\cS,\vx)$.  The function $\Theta(\cS,\vx)$ is the "linking
number" of the surface $\cS$ and the point $\vx$:

\beq
  \Theta\left(\cS,\vx \right) = \cases{1& if $x$ lies inside
  volume bounded by $\cS$; \cr 0& otherwise}\,. \label{Theta}
\eeq
It is obvious that the commutation relation \eq{FPh} is also a consequence
of the Aharonov--Bohm effect.

Now consider the composite operator
\beq
  H_{MN}(\vx,\cS) = \Phi^c_M(\vx)\,
  \cF_N\left(\cS\right)\,,\label{H}
\eeq
where the surface $\cS$ lies at the same time slice as the
point $\vx$.
Using commutation relation \eq{FPh} it is easy to find that:

\beqn
   H_{M_1,N}\left(\vx_1,\cS_1 \right) H_{M_2,N}\left(\vx_2,\cS_2
   \right) - H_{M_2,N}\left(\vx_2,\cS_2\right) H_{M_1,N}
   \left(\vx_1,\cS_1 \right) e^{i \zeta_{12}}\,= 0 \, ,
   \label{comm:H}
\eeqn
where $\zeta_{12} = \frac{2 \pi M_1}{N}
\Theta\left(\vx_1, \cS_2\right) - \frac{2 \pi M_2}{N} \Theta
\left(\vx_2, \cS_1\right)$.
If the point $\vx_1$ lies in the volume bounded by $\cS_2$, the point
$\vx_2$ lies out of the volume $\cS_1$, $M_{1,2}=1$ and \ $N=2$, then
eq.\eq{comm:H} leads to the fermion--like commutation relation

\beq
  H(\vx_1) H(\vx_2) + H(\vx_2) H(\vx_1) = 0\, , \label{HH}
\eeq
where $H(\vx_i) = H_{M_i}(\vx_i, \cS_i)$.

The commutation relations \eq{comm:H} and \eq{HH} can be explained as
follows. The operator $\cF_N(\cS)$ creates the closed world sheet of the
ANO string and the configuration space of the (charged) particles becomes
not simply connected. Similar reasons lead to nontrivial
statistics in $2+1$ dimensions \cite{Rao92}. Note that all operators and
commutation relations considered in the present section can be constructed
in the free theory, but the states created by the operators $U_{str}(\cC_1)$
and $\cF_N(\cS)$ are very unstable in this case. In the Abelian Higgs
theory, the ANO strings exist as a solution of the classical equations of
motion, and this fact justifies the study of the commutation relations
which contain string creation operators.

\section*{Acknowledgments} The authors are grateful to A.S.~Losev,
M.~Minchev, A.D.~Mironov, A.Yu.~Morozov and M.~de~Wild~Propitius for
stimulating discussions.  The authors thank P.~Orland for the useful
comment on the equation \eq{dd}.  This work was supported by the Grant No.
MJM000, financed by the International Science Foundation, by the Grant No.
MJM300, financed by the International Science Foundation and by the
Government of the Russian Federation, by the JSPS Program on Japan -- FSU
scientists collaboration, by the Grant INTAS-94-0840 and by the Grant No.
93-02-03609, financed by the Russian Foundation for Fundamental Sciences.

\newpage

\Appendix{}

Below we show how to derive the expression \eq{YY} for
$J(\tilde x)$. We start from the following definition of $J(\tilde x)$
(see eqs. \eq{def} and \eq{kk}):

\beqn
{[J(\tilde x)]}^{-1}  = \int
\dddd \tilde y_{\mu} \cdot
\delta \left\{ \Sigma_{\mu \nu}
(x, \tilde y) - \Sigma_{\mu \nu}(x, \tilde x) \right\},
\label{JJJ}
\eeqn
where $\Sigma_{\mu \nu}(x, \tilde x) $ and $\Sigma_{\mu \nu}(x, \tilde y)$
are defined by (\ref{theta}).

First we represent functional $\delta$--function in \eq{JJJ} as:

\beqn
\delta \left\{ \Sigma_{\mu \nu} (x, \tilde y) - \Sigma_{\mu \nu}(x,
\tilde x) \right\} =  \nonumber \\
const. \cdot \int \cD k_{\mu\nu}(x)\,\exp\left\{i \int \dd^4 x
k_{\mu\nu}\left(\Sigma_{\mu\nu}(x, \tilde y) - \Sigma_{\mu\nu}(x,
\tilde x)\right)
\right\} = \nonumber \\
const. \cdot \int \dddd k_{\mu \nu}(x) \,
\exp\left\{ i \int_{\Sigma} \ddd^2 \sigma k_{\mu\nu}[\tilde
x(\sigma)] \sqrt{g} t_{\mu\nu} - i \int_{\Sigma'} \ddd^2 \sigma'
k_{\mu\nu}[\tilde y(\sigma')] \sqrt{h} \tau_{\mu\nu}
\right\}, \label{dell}
\eeqn
where $g$ and $t_{\mu\nu}$ are the same as in \eq{theta},
 $h_{ab} = \diff_a \tilde y_{\mu} \diff_b \tilde y_{\mu}$, $h =
det||h_{ab}||$ and $\tau_{\mu\nu} = \frac{\epsilon_{ab}}{\sqrt{h}} \diff_a
\tilde y_{\mu} \diff_b \tilde y_{\nu}$.
Functional integral over $k_{\mu\nu}$ leads to:

\beqn
{[J(\tilde x)]}^{-1} = const \cdot \int \cD \tilde y_{\mu}
\prod_{\mu<\nu}\delta\left(\sqrt{h}\tau_{\mu\nu} -
\sqrt{g}t_{\mu\nu}\right)\,. \label{ID}
\eeqn
Consider now the following functional integral:

\beqn
I(\tilde x) = \cdot \int \cD \tilde
y_{\mu} \cdot \Delta \cdot \delta\left( h^3 - g^3 \right)
\prod_{\mu<\nu}\delta( \tau_{\mu\nu} - t_{\mu\nu}) \nonumber = \\ const
\cdot \int \cD \tilde y_{\mu} \cdot \Delta \cdot \delta\left( 1 -
{\left( \frac{g}{h} \right)}^3 \right)
\prod_{\mu<\nu}\delta(\sqrt{h}\tau_{\mu\nu} - \sqrt{g}t_{\mu\nu})\,.
\label{I:tilde:x2}
\eeqn
Due to the second $\delta$--function, $ \sqrt{h} \tau_{\mu\nu} = \sqrt{g}
t_{\mu\nu}$, and we should assume some regularization\footnote{For example
$\delta^{reg}(x) = \frac{M}{\sqrt{2\pi}}\exp\{-M^2|x|^2\}$, $M
\rightarrow\infty$.} of the first $\delta$--function:  $\delta\left(1 -
{\left( \frac{g}{h} \right)}^3 \right) = \delta^{reg}(0)$. The next
transformations can be accurately performed in the discreetized
space\footnote{The analogous trick was used in \cite{Mor90}.}, but we simply
introduce the parameter $\Lambda$ which plays the role of the inverse
thickness of the string or the ultraviolet cut--off.

\beqn
I(\tilde x) = \int \cD \tilde y_{\mu}
\cdot \Delta \cdot\left(\, \prod_{x\in\Sigma} \delta^{reg}(0)\, \right)
\prod_{\mu<\nu}\delta(\sqrt{h}\tau_{\mu\nu} - \sqrt{g}t_{\mu\nu})
\nonumber \\ = const \cdot \int \cD \tilde y_{\mu}
\cdot \Delta \cdot \exp\left\{\mu S(\Sigma) \right\}
\prod_{\mu<\nu}\delta(\sqrt{h}\tau_{\mu\nu} - \sqrt{g} t_{\mu\nu}) \,,
\label{I:tilde:x1}
\label{Reg}
\eeqn
where $\mu = \Lambda^2 \ln\left( \delta^{reg}(0) \right)$, $S(\Sigma) =
\int_{\Sigma} \ddd^2 \sigma' \sqrt{h(\sigma')}$. The term
$\exp\left\{\mu S(\Sigma) \right\}$ in \eq{I:tilde:x1} is due to the
infinite product of $\delta^{reg}(0)$ over all the points on the
surface $\Sigma$. If we now set

\beqn
\Delta = const \cdot \exp\left\{- \mu S(\Sigma) \right\}\,,
\eeqn
then $I(\tilde x) = {[J(\tilde x)]}^{-1}$,
and

\beq
{[J(\tilde x)]}^{-1} = const \cdot \int \cD \tilde
y_{\mu} \cdot \exp\left\{- \mu S(\Sigma) \right\}\, \delta\left( h^3 - g^3
\right) \prod_{\mu<\nu}\delta( \tau_{\mu\nu} - t_{\mu\nu})\,.
\label{Reg2}
\eeq

The transformations \eq{I:tilde:x2} and \eq{I:tilde:x1} seem to be not very
strict: we have to use $\delta^{reg}(0)$ and the regularization parameter
$\Lambda$. A more accurate derivation of \eq{Reg2} can be done if we notice
that \eq{JJJ} and \eq{dell} is the theory of the Kolb--Ramond field
$k_{\mu\nu}(\tilde y)$, which interacts with the Nambu--Goto string, the
bare string tension being equal to zero. It is important that \eq{Reg2} is
the string theory in which, as we show, the conformal anomaly naturally
arises. This conformal anomaly cancels the conformal anomaly of the original
theory \eq{zzz}, \eq{ii}.

Substituting into \eq{Reg2} the unity of the form:

\beq
1 = \int \dddd h_{ab} \cdot \delta \left( h_{ab} - \diff_a
\tilde y_{\mu} \diff_b \tilde y_{\mu} \right) \dddd \tau_{\mu \nu}
\cdot \delta \left(\tau_{\mu \nu} -
\frac{\epsilon^{ab}}{\sqrt{h}} \diff_a
\tilde y_{\mu} \diff_b \tilde y_{\nu} \right) \label{nn}
\eeq
we obtain:

\beqn
{[J(\tilde x)]}^{-1} = const.\cdot \int \dddd \tilde y \cdot
\dddd h_{ab} \cdot \delta \left( h_{ab} - \diff_a
\tilde y_{\mu} \diff_b \tilde y_{\mu} \right) \dddd \tau_{\mu \nu}
\nonumber \\
\cdot \delta \left(\tau_{\mu \nu} -
\frac{\epsilon^{ab}}{\sqrt{h}} \diff_a
\tilde y_{\mu} \diff_b \tilde y_{\nu} \right) \cdot
\delta (g^3 - h^3) \cdot \delta (t_{\mu\nu} - \tau_{\mu\nu})e^{ - \mu
\int_{\Sigma} \dd^2 \sigma' \sqrt{h}},
\eeqn
It is possible to make the following transformation:

\beqn
\int \dddd h_{a b} \cdot \delta (h_{a b} - \diff_a \tilde y_{\mu} \diff_b
\tilde y_{\mu}) \dddd \tau_{\mu \nu} \cdot \delta \left( \tau_{\mu \nu} -
\frac{\epsilon_{a b}}{\sqrt{h}} \diff_{a}
\tilde y_{\mu} \diff_{b} \tilde y_{\nu} \right) \cdot \nonumber\\
\exp\left\{ - \mu \int_{\Sigma} \ddd^2 \sigma' \sqrt{h(\sigma')}
\right\}...= const \cdot \int \dddd h_{ab} \dddd \tau_{\mu \nu} \nonumber\\
\cdot \exp \left\{ - \int_{\Sigma} \ddd^2 \sigma'\left[\frac{\mu_1}{2}
\sqrt{h} \tau^2_{\mu \nu} + \mu_2 \sqrt{h} h^{a b} \diff_a \tilde y_{\mu}
\diff_b \tilde y_{\mu} + \mu_3 \tau_{\mu \nu} \epsilon_{a b} \diff_a \tilde
y_{\mu} \diff_b \tilde y_{\nu}\right] \right\} \dots  \label{dd}
\eeqn
The fields $h_{ab}$ and $\tau_{\mu \nu}$ have no kinetic terms in the
action, and therefore they acquire their classical values, (see
\cite{Pol81}, or chapter 9 of \cite{Pol87}). If surface $\Sigma$ has
spherical topology we can fix the conformal gauge globally on the surface.
To this end, we substitute the following unity into the integral:

\beq
        1 = \int \cD f \delta(h^f_{12})\,
        \delta(h^f_{11} - h^f_{22}) \cdot \Delta_{FP} \label{Dconf}
\eeq
where $\cD f$ means integration over all possible
reparametrizations, and $h^f_{ab}$ is the change of the metric under a
particular reparametrization $f$. In the last formula, the Faddeev--Popov
determinant $\Delta_{FP}$ appears, which is the exponent of the Liouville
action in conformal gauge with the central charge $- 26$ (see
\cite{Pol87}). The next step is the integration over $\tilde y_{\mu}$.
The first two terms in the exponent (\ref{dd}) give the Liouville action with
the central charge equal to $D = 4$ (see \cite{Pol87}).  And the third term
leads to the term $\frac{\mu^2_3\ln(\Lambda\bar R)}{\mu^2_2 \pi}
\int_{\Sigma} \ddd^2 \tau \sqrt{h} (\diff_a \tau_{\mu \nu})^2$ in the
Jacobian. The integration over $h$ and $\tau_{\mu \nu}$ leads to the
expression (\ref{YY}).

Note that all transformations can be performed in an arbitrary gauge, if
instead of \eq{Dconf} we use the gauge fixing term in the form:

\beq
1 = \int \cD f \prod_{a = b} \delta(h^f_{ab} - g_{ab}) \cdot
\Delta_{FP}\,,
\eeq
here $\Delta_{FP}$ is the exponent of the Liouville action in the
considered gauge.

\end{document}